\documentclass[12pt]{article}
\usepackage{amsmath,amssymb,epsfig}
\usepackage{cite}

\usepackage{color}
\input{colordvi.tex}
\def\unit{{\relax{\rm 1\kern-.26em I}}}

\makeatletter

\renewcommand\section{\@startsection {section}{1}{\z@}%
                                  {-3.5ex \@plus -1ex \@minus -.2ex}%
                                  {2.3ex \@plus.2ex}%
                                  {\normalfont\large\bfseries}}

\renewcommand\subsection{\@startsection{subsection}{2}{\z@}%
                                    {-3.25ex\@plus -1ex \@minus -.2ex}%
                                    {1.5ex \@plus .2ex}%
                                    {\normalfont\normalsize\bfseries}}

\newcount\hour \newcount\minute
\hour=\time \divide \hour by 60
\minute=\time
\def\now{%
\ifnum \hour<13
 \ifnum \hour=0 \advance \hour by 12 \number\hour:\else \number\hour:\fi%
    \ifnum \minute<10 0\fi%
    \number\minute%
\ A.M.%
\else \advance \hour by -12 \number\hour:%
 \ifnum \minute<10 0\fi%
 \number\minute%
 \ P.M.%
\fi%
}

\makeatother

\begin{document}

\baselineskip=18pt  
\numberwithin{equation}{section}  
\allowdisplaybreaks  



%
%


\thispagestyle{empty}

\vspace*{-2cm}
\begin{flushright}
\end{flushright}

\begin{flushright}
YGHP-12-53\\
KUNS-2427\\
DESY 12-219 \\
OCU-378
\end{flushright}

\begin{center}

\vspace{1.4cm}

\vspace{0.5cm}
{\bf \Large Cosmic R-string, R-tube and Vacuum Instability}
\vspace*{0.5cm}

{\bf
Minoru Eto$^{1}$, Yuta Hamada$^{2}$, Kohei Kamada$^{3}$, \\Tatsuo
Kobayashi$^{2}$, Keisuke Ohashi$^{4}$ and Yutaka Ookouchi$^{2,5}$}
\vspace*{0.5cm}

\vspace*{0.5cm}

$^{1}${\it {Department of Physics, Yamagata University, Yamagata 990-8560, Japan}}
\vspace{0.1cm}

$^{2}${\it Department of Physics, Kyoto University, Kyoto 606-8502, Japan}\\

\vspace{0.1cm}

$^3$ {\it Deutsches Elektronen-Synchrotron DESY,
Notkestrasse 85, D-22607 Hamburg, Germany }\\

\vspace{0.1cm}

$^{4}${\it Department of Mathematics and Physics, Osaka City
 University, Osaka 558-8585, Japan  }\\

\vspace{0.1cm}

$^{5}${\it The Hakubi Center for Advanced Research, Kyoto University, Kyoto 606-8302, Japan }\\

\vspace*{0.5cm}

\end{center}

\vspace{1cm} \centerline{\bf Abstract} \vspace*{0.5cm}

We show that a cosmic string associated with spontaneous $U(1)_R$ symmetry 
breaking gives a constraint for supersymmetric model building. In some models, 
the string can be viewed as a tube-like domain wall with a winding number interpolating a false 
vacuum and a true vacuum. Such string causes 
inhomogeneous decay of the false vacuum to the true vacuum via rapid 
expansion of the radius of the tube and hence its formation would be 
inconsistent with the present Universe. However, we demonstrate that there exist metastable solutions 
which do
not expand rapidly. Furthermore, when the true vacua are degenerate, the structure inside the tube becomes involved. As an example, we show a ``bamboo''-like solution, which suggests a possibility observing an information of true vacua from outside of the tube through the shape and the tension 
of the tube.  

\newpage
\setcounter{page}{1} 



\section{Introduction}

The global $U(1)_R$ symmetry plays an important role 
in supersymmetric field theories, in particular in 
supersymmetry (SUSY) breaking  
\cite{Nelson:1993nf,Intriligator:2007py,KOO,Abe:2007ax,Kang:2012fn} (See \cite{rev1,rev2,rev3} for reviews and references therein). In \cite{KOO,MurayamaNomura,Intriligator:2007py}, by exploiting the Nelson-Seiberg theorem \cite{Nelson:1993nf}, a connection between metastability and R-symmetry was demonstrated in the context of generalized Wess-Zumino models with generic superpotential.
{}From more phenomenological viewpoint, the $U(1)_R$ symmetry 
must be broken explicitly or spontaneously to generate 
Majorana gaugino masses.
Gaugino masses are not induced by SUSY breaking without 
the $U(1)_R$ symmetry breaking.

Indeed, several types of models for the $U(1)_R$ symmetry breaking 
have been studied
\cite{Kitano1,Kitano2,Shih:2007av,Ferretti:2007ec,Cho:2007yn,Abel:2007jx,
Aldrovandi:2008sc,Carpenter:2008wi,Giveon:2008ne,Sun:2008va,Komargodski:2009jf,Azeyanagi:2012pc,R1,R2,Stone}.
In some models, the vacuum with both SUSY and $U(1)_R$ breaking 
may be a global minimum.
However, in many models such vacuum is 
a metastable minimum and there is a global minimum, 
where SUSY and $U(1)_R$ may be unbroken.

Through the cosmological phase transition, there may appear 
solitonic objects 
such as domain walls, cosmic strings and monopoles
\cite{Kibble:1980mv} through the Kibble-Zurek mechanism \cite{Kibble,Zurek}.
When a global $U(1)$ symmetry is spontaneously broken, 
a global string appears \cite{Vilenkin}.
Thus, when the $U(1)_R$ symmetry is broken spontaneously in SUSY models,  
there would appear a global string, which we refer as  
an R-string.\footnote{See also for another type of strings, which 
appear through SUSY breaking \cite{Eto:2006yv,Hanaki}.}
The R-string would be stable in those models in which 
the $U(1)_R$ breaking vacuum is a global minimum, 
and that would lead to several cosmologically interesting aspects \cite{OurII}.

On the other hand,  when the $U(1)_R$ breaking vacuum is 
metastable and the model has another global minimum with 
SUSY and $U(1)_R$ unbroken, 
there may appear an R-string, whose core corresponds to 
the true SUSY vacuum, i.e., R-tube. One may think that such an R-tube is unstable because 
the energy density in the core, which is the SUSY vacuum, 
is lower than one outside, which is the SUSY breaking metastable vacuum. Thus, it would ``roll-over'' and the true SUSY 
vacuum would expand in the Universe \cite{Steinhardt:1981ec,Hosotani,Yajnik:1986tg}. 
In this case, the SUSY breaking could not be realized successfully.
One may conclude that a scenario with R-tube formation is ruled out
by this mechanism.

However, since the domain wall tension works as a centripetal force
for R-tube, its radius may be stabilized if the domain wall tension is 
large enough and the energy discrepancy between SUSY vacuum and SUSY-breaking
vacuum is small enough. 
In such a case, the cosmological disaster can be avoided. 
The (in)stability of the R-tube soliton depends on 
parameters in the SUSY models.
In principle, we can have constraints on SUSY-breaking models 
from this consideraion 
because (in)stability of the R-tube is determined by 
parameters of SUSY breaking models. Note that such constraints are independent of 
the requirement that the metastable vacuum decays slowly into 
the true vacuum by the tunneling effect \cite{Coleman:1977py}, 
compared with the Universe age.
Therefore, it is quite important to study the R-string/R-tube 
formation and its (in)stability.
Some relevant studies have been carried out in
Refs.~\cite{Kumar:2008jb,Kumar:2009pr}.

In this paper, we study in detail the structure of 
the R-string/tube solution in SUSY models.
In a simple but (semi)realistic SUSY breaking model, 
we study stability of the R-tube by exploiting a piecewise linear 
approximation and numerical solutions. 
By using linear approximation, we obtain constraints for the 
stable R-tube. We also show examples of (meta)stable/unstable 
R-tube configurations numerically. 
We emphasize that the winding number, which is an important 
quantity to characterize features of the R-tube solutions, 
is also relevant to the stability of the R-tube.

We also show that the core of the R-tube can have more 
complicated structure in certain SUSY models where 
the true SUSY breaking vacua are degenerate.
For example, suppose that the SUSY model has a $\mathbb{Z}_2$ symmetry 
and it is broken at the true SUSY vacuum.
Then, the core of the R-tube would be separated into 
two vacua by a domain wall.
Since it looks like a (gourd-shaped) bamboo, 
we refer it as the bamboo solution.
We also study aspects of the bamboo solution.
Other types of structure inside the core of strings would be 
possible.

This paper is organized as follows.
In section 2, we illustrate the R-string and tube solutions 
in simple models as warm-up.
In section 3, we study the R-tube 
in a (semi)realistic but simple SUSY breaking model, that is, 
an O'Raifeartaigh-like model with non-canonical K\"ahler metric.
We analyze the (in)stability of the R-tube 
numerically at several parameters.
In section 4, by showing the bamboo solution, we demonstrate the fact that quantum number in the SUSY vacua significantly affects the shape and the tension of the string.
Section 5 is devoted to conclusion and discussion.
In appendix A, we show basics of the relaxation method for solving a differential equation.

\section{Stable R-string and tube solutions}

Before going to detailed studies of metastable strings, which will be shown in the next section, we would like to present simple stable solutions as a warm-up. Here, we illustrate a stable R-string and a stable R-tube which is a tube-like domain wall with winding number, by using single
complex scalar field models.

\subsection{R-string}
\label{sec:R-string}


Consider the following simplest spontaneous R-symmetry breaking model. Superpotential is linear in a chiral superfield 
$X$ which will be a trigger for SUSY breaking,
$${
W=f X. 
}$$
To stabilize the pseudo-moduli $X$ in this SUSY breaking vacuum, we introduce the following non-canonical K\"ahler potential by hand, 
\begin{eqnarray}
g_{X \bar X}^{-1} = 1 - \frac {\mu_X^2}{f^2}|X|^2 + \frac{\lambda_X}{4f^2}|X|^4.
\end{eqnarray}
Thus, the potential of this theory is given by
\begin{eqnarray}
V(X) = f^2 - \mu^2_X |X|^2 + \frac{\lambda_X}{4}|X|^4. \label{Shihex1}
\end{eqnarray}

This model can be viewed as a low energy effective theory of one of
the O'Raifeartaigh models studied in \cite{Shih:2007av}: When the
pseudo-moduli space is stable everywhere along messenger directions,
by integrating out the messengers, one obtains non-trivial corrections
to the K\"ahler potential. Expanding the K\"ahler potential up to
${\cal O}(|X|^6)$, one can reproduce a theory similar to \eqref{Shihex1}. On the other hand, when the pseudo-moduli space has a tachyonic direction at a point in the space, which is phenomenologically interesting situation in gauge mediation models \cite{Komargodski:2009jf,KOO}, the existence of messengers is crucial and two-field model is required. This is the main topic in the next section.


When $\mu_X^2 > 0$ and $\lambda_X >0$, 
the $X$ field develops its vacuum expectation value and the R-symmetry is broken ($X$ has the R charge 2).
The minimum of the potential $V(X)$ is obtained at 
$|X|=X_{\rm min} \equiv \sqrt{2\mu_X^2/\lambda_X}$.
Note that this vacuum is the global minimum of the potential $V(X)$.
Since the global $U(1)_R$ symmetry is spontaneously broken, 
the R-string would be formed.

\begin{figure}
\begin{center}
\includegraphics[width=16cm]{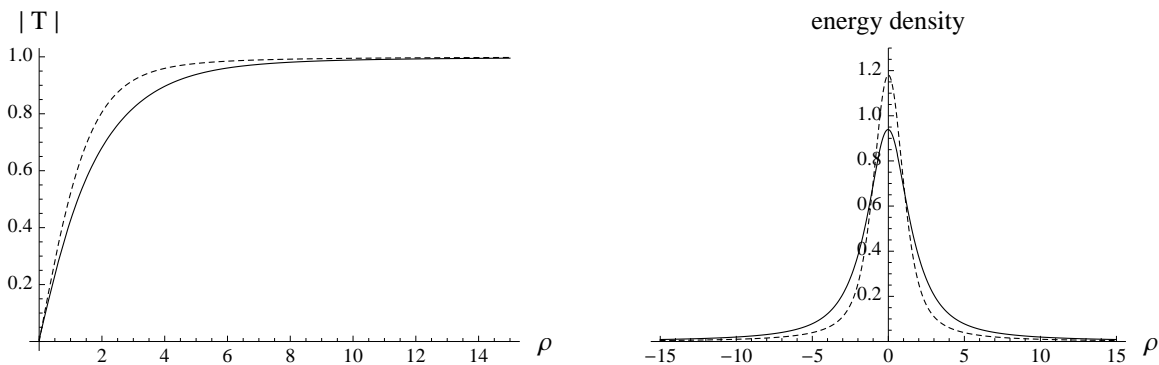}
\caption{\sl The minimal winding solution of the R-string (solid line) for 
$\lambda = 1/2$: 
the profile function is shown in the left panel 
and the corresponding energy density is shown in the right panel.
The broken lines stand for the standard global vortex which is the solution of the model with the minimal
kinetic term in Eq.~(\ref{eq:lag_sol1}).}
\label{fig:sol1}
\end{center}
\end{figure}

Let us introduce dimensionless variables as
\begin{eqnarray}
X = X_{\rm min} T,\quad
x_\mu = \sqrt{\frac{2\mu_X^2}{\lambda_X f^2}}\ \tilde x_\mu,\quad
\lambda = \frac{\lambda_X f^2}{4\mu_X^4}>0.
\end{eqnarray}
Then the effective Lagrangian is given by
\begin{eqnarray}
{\cal L} = f^2
\left[ \frac{\left| \tilde \partial_\mu T\right|^2}{{\cal V}(T)} - 
{\cal V}(T)\right], \quad {\cal V}(T)\equiv 1-\frac{|T|^2}{2\lambda}+\frac{|T|^4}{4\lambda}.
\label{eq:lag_sol1}
\end{eqnarray}
Here a positive definite metric at the minimum $(T=1)$ requires $\lambda
>1/4$.
We use the dimensionless cylindrical coordinate $(\rho,\theta,\tilde z)$ for constructing a straight R-string along the $z$-axis.
We make the following standard Ansatz,
\begin{eqnarray}
T(\rho,\theta,\tilde z) =f(\rho)e^{i n\theta}, 
\end{eqnarray}
where $f(0) =0$ and $f(\rho) \rightarrow 1$ at $\rho \rightarrow \infty$.
We numerically solve the equation of motion for a minimal winding solution ($n=1$).
The solution is shown in Fig.~\ref{fig:sol1}.

For later convenience, let us estimate a size of R-string, $R$,
by using the following simple approximation
\begin{eqnarray}
 f(\rho)=\left(\frac{\rho}{R}\right)^n {\rm for~} \rho \le R 
\quad {\rm and~} \quad  f(\rho)=1  \quad {\rm for~} \rho >R, \label{eq:approxf}
\end{eqnarray} 
where the power of $\rho$ is determined by requiring smoothness of
the configuration at $\rho=0$. The total energy of this configuration, $E$, 
per the string length $\Delta z$ is
estimated as  
\begin{eqnarray}
 \frac{\lambda_X}{2\mu_X^2}\frac{E}{2\pi \Delta z}\approx c_n(\lambda)+\frac{n^2}{{\cal V}(1)} \log
  \left(\frac{\Lambda}{R}\right)
+\frac12 {\cal V}(1) \Lambda^2+\frac12 \frac{a_n}{4\lambda} R^2,\quad 
 a_n=\frac{2n^2}{1+3n+2n^2},
\end{eqnarray}
where  an $R$ independent constant
 $c_n(\lambda)$ which should be numerically determined is introduced, and an IR-cutoff $\Lambda$ is 
also introduced in order to regularize a 
well-known logarithmic divergence of a global vortex.
The above energy takes the minimum at $ R\approx R_{\rm string}$,
\begin{eqnarray}
 R_{\rm string}\equiv \frac{2n}{\sqrt{a_n} m_T}\approx \frac{2n }{m_T},\qquad m_T^2
\equiv \frac1\lambda \left(1-\frac1{4\lambda}\right),\label{eq:RstringSize1}
\end{eqnarray}
which is the transverse size of the R-string.
Here $m_T$ is the dimensionless mass of $T$ in the vacuum $T=1$.
For instance for $\lambda=1/2$ and $n=1$, we obtain $m_T=1$ and 
$R_{\rm string}=2\sqrt{3}$.

\begin{figure}
\begin{center}
\includegraphics[width=8cm]{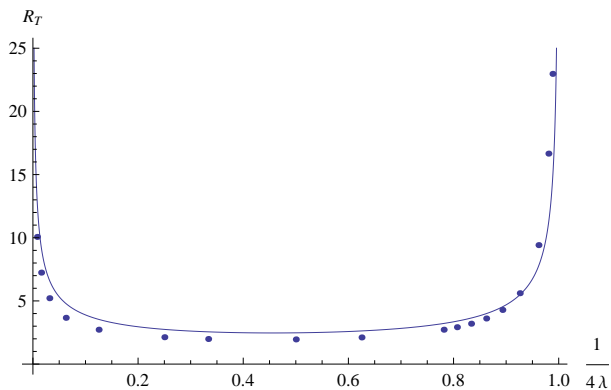}\quad
\caption{\sl The transverse sizes of the R-strings, $R_T$, with the linear approximation (solid line)
 and numerical calculations (points) using the relaxation method
 reviewed in Appendix \ref{sec:relaxation}.
}
\label{fig:RStringSize}
\end{center}
\end{figure}

To check the approximation (\ref{eq:approxf}) 
by comparing with numerical calculations, 
we introduce another definition of 
a transverse size of the R-string  as
\begin{eqnarray}
 R_T \equiv 
\frac{\int_0^\infty d\rho\,  \rho^2\, K_T}
{\int_0^\infty d\rho\,  \rho\, K_T},\quad K_T=
\frac{(f'(\rho))^2}{{\cal V}(f(\rho))}.\label{eq:RstringSize2}
\end{eqnarray}
Here  $K_T$ gives a finite contribution from the kinetic term along the $\rho$
direction into the energy and is useful to define the transverse size.
Note that this quantity does not include the cut-off dependence.
We compute this both analytically with the linear approximation and numerically, see Fig.\ref{fig:RStringSize}.
For instance 
we observe $R_T=1.95$ for $\lambda=1/2$ by a numerical calculation.
As can be seen in Fig.~\ref{fig:RStringSize}, the linear approximation nicely reproduces the 
numerical results
(we need to pay attention to errors of 10\% $-$ 30\%  in the linear approximation).

\subsection{Tube solution}
\label{sec:Tube}

\begin{figure}
\begin{center}
\includegraphics[width=16cm]{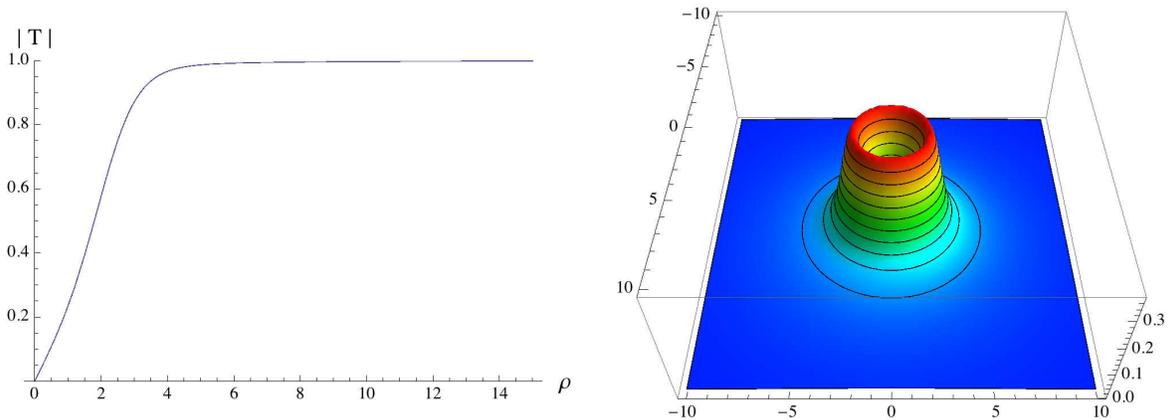}
\caption{\sl The minimal winding solution of the string of the model 
given in Eq.~(\ref{eq:model_sol2}). 
The profile function is shown in the left panel 
and a slice at a fixed $z$ of the corresponding energy density is shown in the right panel.}
\label{fig:sol2}
\end{center}
\end{figure}

Here, in order to illustrate the tube solution, we 
study \emph{a non-supersymmetric bosonic theory} as a toy model.
Let us study the model with the following scalar potential, 
\begin{eqnarray}
V(X) = \frac{|X|^2}{M^2}(|X|^2 - v^2)^2,
\end{eqnarray}
and the canonical kinetic term, 
where $M$ and $v$ are taken to be real.
This model has a global $U(1)$ symmetry (no longer $U(1)_R$ symmetry), under which 
the field $X$ transforms.
This potential has two degenerate vacua, that is, 
$|X|=0$ and $v$.
At the former vacuum, the global $U(1)$ symmetry is unbroken, 
while the $U(1)$ symmetry is broken at the latter vacuum. 
Then, a global string would be formed.
Again, let us rescale the fields and coordinates as
\begin{eqnarray}
X = v T,\quad x_\mu =  \frac{M}{v^2}\tilde x_\mu,
\end{eqnarray}
then the Lagrangian becomes
\begin{eqnarray}
{\cal L} = \frac{v^6}{M^2}\left[\left|\tilde\partial_\mu T\right|^2 - |T|^2\left(1-|T|^2\right)^2\right].
\label{eq:model_sol2}
\end{eqnarray}
We make the Ansatz for the minimally winding string,
\begin{eqnarray}
T=f(r)e^{i n\theta}, 
\end{eqnarray}
where $f(0) =0$ and $f(r) \rightarrow 1$ at $r \rightarrow \infty$.
The solution is again obtained numerically which is shown in Fig.~\ref{fig:sol2}.
As can be seen in Fig.~\ref{fig:sol2}, the string has a substructure that is a hole inside the string.
Thus, we refer it as the tube.
It is the asymmetric phase ($X\neq0$) outside the tube while it is the symmetric ($X=0$) phase inside it.

This tubelike string solution can be regarded as a tube of a domain wall.
Indeed, there also exists a domain wall in this model. For instance, a  solution
interpolating the two vacua $T=1$ at $x_1=\infty$ and 
$T=0$ at $x_1=-\infty$ is given by
\begin{eqnarray}
 T=\frac1{\sqrt{1+e^{-2 \tilde x_1}}} ,
\end{eqnarray} 
with a dimensionless tension $T_{\rm wall}=1/2$. Thus, assuming that the field configuration of the tube along the radial direction is well described 
by this solution, the total  energy $E$ per the tube length $\Delta z$ of the tubelike solution with a radius $\rho = R$  can be estimated by
\begin{eqnarray}
\frac1{v^2} \frac{E}{2\pi \Delta z}\approx  T_{\rm wall} R+n^2 \log\left(\frac{\Lambda}{R}\right), 
\end{eqnarray}
as long as the ``thickness'' of the wall is much smaller than the radius $R$. 
Minimizing this, we get the transverse size of the tube solution
\begin{eqnarray}
 R\approx \frac{n^2}{T_{\rm wall}}=2n^2.
\end{eqnarray} 
Note that the stabilization mechanism of the tube solution is different from 
that of the R-string (without a hole) 
where the kinetic energy and the potential energy are balanced.
As a result, the transverse sizes have different dependences on
the winding number $n$ as 
$R\propto n$ for the R-strings and $R\propto n^2$ for the tubes,
respectively.   

\begin{figure}
\begin{center}
\includegraphics[width=8cm]{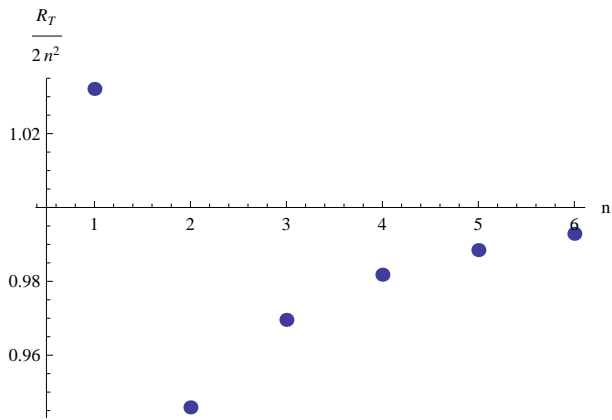}
\caption{\sl Numerical results 
of a ratio $R_T/2n^2$ for higher winding solutions,
by using the relaxation method with efficiently large relaxation time,
$\tau>20\times n^4$.}
\label{fig:RTvsWinding}
\end{center}
\end{figure}

We can define a transverse size $R_T$ of this tube-like string 
similar to Eq.(\ref{eq:RstringSize2}) with $K_T=(f'(\rho))^2$ 
and observe $R_T=2.06$ for the minimal winding tube by a numerical
calculation. Ratios $R_T/2n^2$ with higher winding solutions are listed 
in Fig.\ref{fig:RTvsWinding}. 
It suggests that the above approximation works well. 

\section{R-tube and Vacuum Instability }
\label{R-tubeInstability}

\subsection{Metastable R-tube}


In section \ref{sec:R-string}, we studied the single field SUSY breaking model as a toy model of one of the O'Raifeartaigh models discussed in \cite{Shih:2007av} in which classical pseudo-moduli space is stable everywhere. In this section, we move on to phenomenologically interesting situation where pseudo-moduli space has a tachyonic direction, in which large gaugino masses are generated by gauge mediation \cite{Komargodski:2009jf,Ookouchi}. In such models, the R-symmetry breaking vacuum is metastable, thus the R-string solution can be a tube-like domain wall with winding number as showed in section \ref{sec:Tube}.

Here, we study the illustrating supersymmetric model with 
two superfields, $X$ and $\phi$.
These superfields have R-charges, $R[X]=2$ and $R[\phi]=0$, 
and the superpotential is given by 
\begin{eqnarray}
W= X\phi^2 - \mu^2 X.
\end{eqnarray}
In addition, we consider the following effective K\"ahler metric,
\begin{equation}
g_{X\bar{X}}^{-1}=1-{1\over 2m^2} |X|^2 +{\lambda \over 4 m^4}|X|^4,
\quad g_{\phi \bar{\phi}}^{-1}=1, \quad
g^{-1}_{X\bar{\phi}}=g^{-1}_{\phi \bar{X}}=0 .
\end{equation}
This model has the $\mathbb{Z}_2$ symmetry, under which $X$ and $\phi$ 
are $\mathbb{Z}_2$ even and odd, respectively.
This model has the discrete SUSY vacua, 
\begin{eqnarray}
X=0, \qquad \phi = \pm \mu,
\end{eqnarray}
and the SUSY breaking vacuum,
\begin{eqnarray}
X= \frac{m}{\sqrt{\lambda}}, \qquad \phi =0,
\end{eqnarray}
where the $U(1)_R$ symmetry is also broken.
The former is the true vacuum, while the latter is 
the metastable vacuum whose vacuum energy is 
$V=\mu^4{\cal V}(1)=\mu^4\left(1-\frac{1}{4\lambda}\right)$.

For later convenience, let us introduce dimensionless variables by
\begin{eqnarray}
X = \frac{m}{\sqrt{\lambda}} T,\quad \phi = \mu s,\quad x_\mu = \frac{m}{\sqrt{\lambda}\mu^2}\tilde x_\mu,
\quad \epsilon = \frac{\sqrt{\lambda} \mu}{m},
\end{eqnarray}
then the Lagrangian is of the form
\begin{eqnarray}
{\cal L} &=& \mu^4\left[
\frac1{{\cal V}(T)}
\left|\tilde \partial_\mu T\right|^2
+ \epsilon^2\left|\tilde \partial_\mu s\right|^2 - 
{\cal V}(T)|s^2-1|^2
- \frac{4}{\epsilon^2}|T|^2 |s|^2
\right].
\label{eq:model_sol3}
\end{eqnarray}
This Lagrangian is characterized by two dimensionless parameters
$\lambda$ and $\epsilon$.
For instance, in the SUSY breaking vacuum $(T,s)=(1,0)$, 
dimensionless masses for $T$ and $s$ are wriiten by 
\begin{eqnarray}
 m_T^2=\frac1{\lambda}\left(1-\frac1{4\lambda}\right),\quad 
m_s^2=\frac{2}{\epsilon^2}\left(\frac{2}{\epsilon^2}+\frac1{4\lambda}-1\right),
\end{eqnarray}  
respectively, and that is, existence of the SUSY breaking vacuum requires
\begin{eqnarray}
 1> \frac1{4\lambda}>0,\quad   \frac{2}{\epsilon^2}+\frac1{4\lambda}>1.
\label{eq:VacuumCond}
\end{eqnarray} 

\begin{figure}[t]
\begin{center}
\includegraphics[width=16cm]{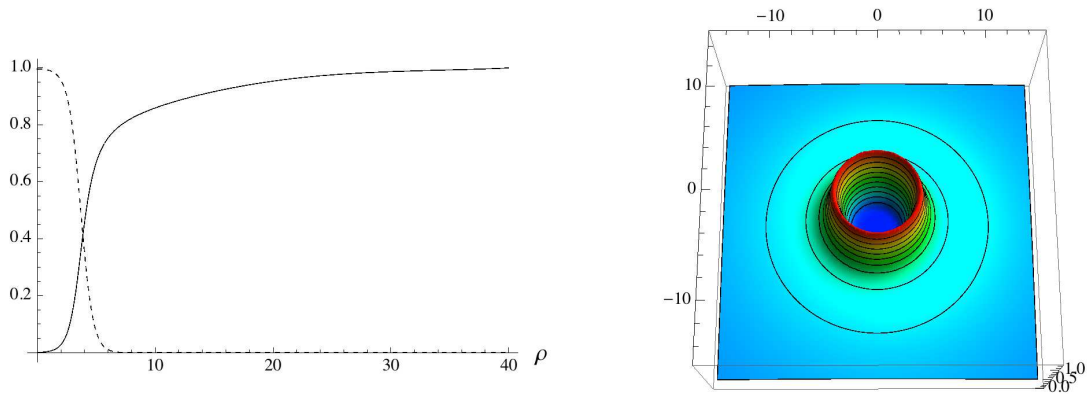}
\caption{\sl The minimal winding solution of the R-tube of the two-scalar model for $\epsilon=1$, $\lambda=0.27$
given in Eq.~(\ref{eq:model_sol3}). 
The profile functions (solid line for $|T|$ and broken line for $s$) are shown in the left panel 
and a slice at a fixed $z$ of the corresponding energy density is shown in the right panel.
Here we chose the plus sign ($\gamma = +1$).}
\label{fig:sol3}
\end{center}
\end{figure}
If one is interested in vacuum selection, a simple criterion is a ratio
of tachyonic masses at the origin $(T,s)=(0,0)$ where we have  
\begin{eqnarray}
 (m_{0,T})^2=-\frac1{2\lambda},\quad (m_{0,s})^2=-\frac{2}{\epsilon^2}.
\end{eqnarray}
Since in the early universe field values are assumed  to be around the
origin, if tachyonic mass of $T$ is larger, one may expect that supersymmetry breaking model is preferable\footnote{Here we have been studying a simple toy model. To discuss vacuum selection more seriously, one need to go back to an original realistic model and specify a history of the early universe.}. An inequality
\begin{eqnarray}
 \frac2{\epsilon^2}<\frac1{2\lambda} , \label{redline}
\end{eqnarray}
is required for selecting the SUSY breaking vacuum.

Now we are ready to construct the R-tube in this two-scalar model.
To this end, we make the Ansatz
\begin{eqnarray}
T = f(\rho)e^{in\theta},\quad 
s = \gamma\, h(\rho),
\label{eq:ansataz_sol3}
\end{eqnarray}
with $\gamma = \pm 1$. Because of the $\mathbb{Z}_2$ symmetry, 
the solutions of $f(\rho),h(\rho)$ are independent of $\gamma$. 
Similar to the model in section~\ref{sec:Tube}, it is the symmetric phase inside the tube while
it is the asymmetric phase outside the tube.  A sharp contrast among two models is that the outside 
is the true vacuum in section~\ref{sec:Tube} and is the false vacuum in this section. One may guess that stable solution does not exist since the core of the tube has 
lower energy than its outside and hence larger radius would be favored energetically, 
which causes the ``roll-over'' problem. 
However, because the tension of the wall acts as a centripetal force for the R-tube, 
we will find that there exist metastable tube-like field configurations. 
In order to see a typical R-tube numerical solution in this model, here we show an example in Fig.~\ref{fig:sol3} with $\epsilon =1, \lambda = 0.27$.
%
%
Note that the profile function of the winding field $T$, whose mass is very small, has a very long tail compared to that 
of the solution in section~\ref{sec:Tube}. On the other hand, the unwinding field $s$, whose dimensionless mass is of order 1, converges exponentially.
In the next subsection, we will investigate stability of the R-tube by varying those parameters.

\subsection{Instability of R-string and Broken $\mathbb Z_2$ Symmetry}

If we set $s=0$ to keep the $\mathbb Z_2$ symmetry, 
the model discussed in this section reduces to 
just the model discussed in section \ref{sec:R-string} except for an overall
factor. Therefore the R-string solution $(s=0)$
without a hole inside  is also a solution in this model. 
However, 
such an R-string would be almost
always unstable and transforms into an R-tube with non-zero $s$ inside.
Since non-vanishing $s$ means  
the broken $\mathbb Z_2$ symmetry, we observe below that 
the $\mathbb Z_2$ symmetry inside the R-tube in this model
is almost always broken.  

\begin{figure}[ht]
\begin{center}
\includegraphics[width=7cm]{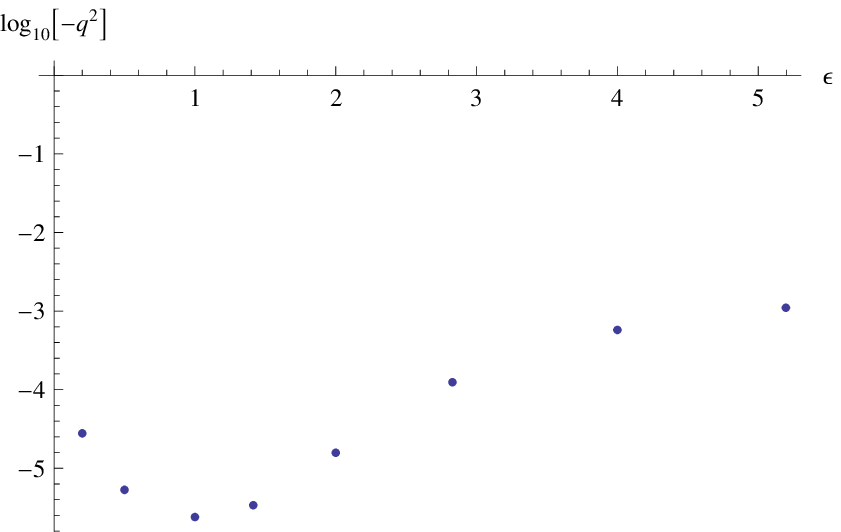}\quad
\includegraphics[width=7cm]{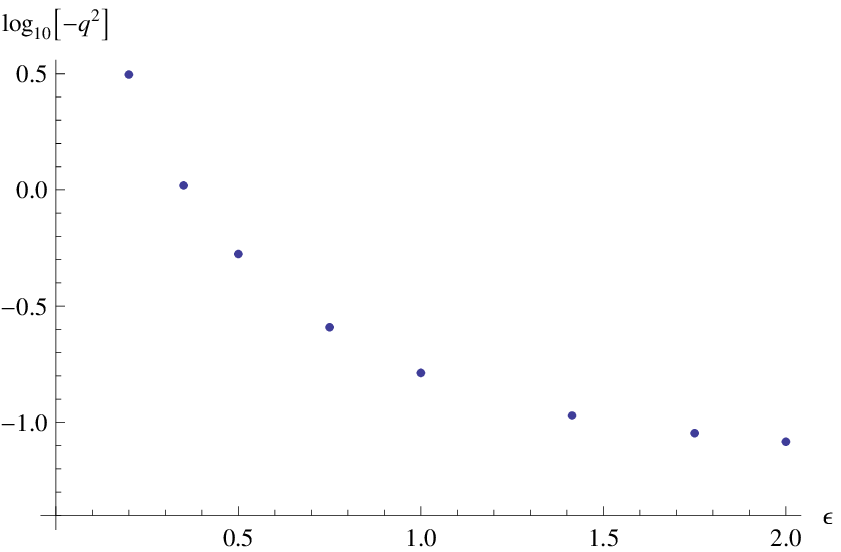}
\caption{\sl Tachyonic mass of $s$ around R-string for $\lambda=0.27$
 (the left panel) and $\lambda=1/2$(the right panel).} 
\label{eq:RstringStability}
\end{center}
\end{figure}

Let us consider 
an infinitesimal fluctuation $s (\ll 1)$ 
around the R-string solution discussed in section \ref{sec:R-string}
and study whether a direction along $s$ is tachyonic or not.
A linearized equation for $s$ is given with an eigenvalue $q^2$ as  
\begin{eqnarray}
 -\frac1\rho (\rho s')'+\left(-\frac{2}{\epsilon^2}{\cal
			 V}(T)+\frac{4}{\epsilon^4}|T|^2\right)
\Big|_{T=T_{\rm  sol}} \,s= q^2 s. \label{eq:tachyonic_mass}
\end{eqnarray}
For instance we observe tachyonic masses of $s$ numerically for many sets
of parameters $\{\lambda,\epsilon\}$ as shown 
in Fig.\ref{eq:RstringStability}. 
We therefore make a conjecture that 
\begin{eqnarray}
 q^2 <0 ,
\end{eqnarray} 
for almost all the winding number $n$ and the almost whole
parameter region of $\{\lambda, \epsilon\}$ satisfying the 
inequalities (\ref{eq:VacuumCond}). 
This conjecture means that the 
R-string with $s=0$ is always unstable and a stable R-tube solution, if it exists, 
must have the following property
\begin{eqnarray}
s|_{\rho=0} \not=0.
\end{eqnarray}
In this paper we will assume this conjecture holds and will not consider the 
constraints from the stability of R-string configuration.

\subsection{Rough Estimation for R-tube}

\begin{figure}[ht]
\begin{center}
\begin{picture}(50,50)(0,0)
\put(70,-10){\footnotesize $R_{\rm tube}$}
\put(150,-10){\footnotesize $R_{\rm max}$}
\end{picture}
\includegraphics[width=7cm]{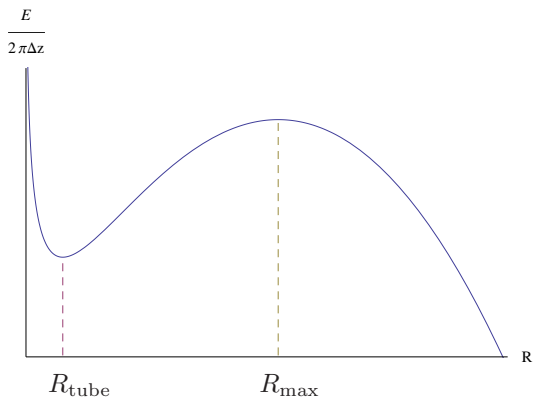}
\caption{\sl Potential for a transverse size of R-tube with 
a rough estimation.} 
\label{fig:ToyPotential}
\end{center}
\end{figure}

If the domain wall consisting the R-tube is sufficiently thin and resides at 
$\rho=R$, 
its total energy $E$  per the length $\Delta z$ can be estimated as   
\begin{eqnarray}
 \frac{\lambda}{m^2}\frac{E}{2\pi \Delta z}&\approx& \frac12 {\cal V}(1) (\Lambda^2-R^2)
+T_{\rm wall} R+\frac{n^2}{{\cal V}(1)}\log
\left(\frac{\Lambda}{R}\right), 
\label{eq:toymodel}
\end{eqnarray}
as has been done in Sec.~\ref{sec:Tube}. 
See Fig.\ref{fig:ToyPotential}. 
Note that 
the total energy has divergence terms with an IR-cutoff $\Lambda$
proportional to
$\Lambda^2$ and $\log \Lambda$.
The former is the energy density ${\cal V}(1)=1-\frac1{4\lambda}$ of the SUSY breaking
vacuum, and the latter is that for the well-known global string tension. 
Here, $T_{\rm wall}$ is a (dimensionless) tension of the domain wall.
This potential has a local minimum (maximum) at $\rho=R_{\rm tube}
(R_{\rm max})$ with  
\begin{eqnarray}
 R_{\rm max}&=& 
 \frac{2n^2}{ {\cal V}(1)\left(T_{\rm wall}-\sqrt{T_{\rm wall}^2-4n^2}\right)}> \frac{n}{{\cal V}(1)},\quad\nonumber \\
R_{\rm tube}&=&\frac{2n^2}{{\cal V}(1)(T_{\rm wall}+\sqrt{T_{\rm
 wall}^2-4n^2})}<\frac{n}{{\cal V}(1)},
\end{eqnarray}
if the dimensionless tension $T_{\rm wall}$ is sufficiently large as 
\begin{eqnarray}
 T_{\rm wall}> 2n.\label{eq:stability1}
\end{eqnarray}  
This is, therefore, a necessary and sufficient condition 
for existence of the R-tube 
as long as the approximation Eq. (\ref{eq:toymodel}) is valid. 
Such configurations that satisfy the inequality
$R\ge R_{\rm max}$ can not avoid to spread
out toward the infinite of the space.  
Note that comparing a thickness $L_{\rm wall}$ of the domain wall with $R_{\rm tube}$,  
if $L_{\rm wall}\ll R_{\rm tube}$ holds,
the above estimation (\ref{eq:toymodel}) works well  and we will observe the SUSY vacuum inside the R-tube, namely
\begin{eqnarray}
s|_{\rho=0}  \approx 1.
\end{eqnarray}
Using an approximation discussed in the next subsection, we can show  
the lower limit of the ratio 
\begin{eqnarray}
\frac{R_{\rm tube}}{L_{\rm wall}} >\frac{n^2}{{\cal V}(1) T_{\rm
 wall}L_{\rm wall} } > \frac{n^2}{2(1+\epsilon^2 {\cal
 V}(1))}> \frac{n^2}6.
 \label{eq:bound}
\end{eqnarray}
Therefore $R_{\rm tube}$ can not be very small. If $R_{\rm tube}$ is
comparable with $L_{\rm wall}$, a configuration of R-tube approaches
one of the R-string, but $s|_{\rho=0} $ keeps non-vanishing even there 
in almost all the cases as we discussed.
%
%
%
%
\subsection{Linear Approximation for the Domain Wall \label{Linear}}

In order to estimate the transverse size of the R-tube and its stability 
following the discussion in the previous subsection, we need the data $\{T_{\rm wall},L_{\rm wall}\}$. 
We here evaluate them assuming that 
the domain wall in the R-tube can be well approximated by a flat domain
wall interpolating the SUSY vacuum at $x=-\infty$ and the SUSY
breaking vacuum at $x=\infty$.  Let us consider this configuration in
the following. 
Note that, however, 
 there is an ambiguity for definition of $T_{\rm wall}$ and 
profile functions for the domain wall 
since the flat domain wall itself is unstable.  
It is natural to set a relation between the total energy
of the system $E_{\rm wall}$ and the tension $T_{\rm wall}$ of the domain
wall sitting at $x=\left<x\right>$ with IR-cutoff $\Lambda_\pm$  as
\begin{eqnarray}
 E_{\rm wall}=\int_{\Lambda_-}^{\Lambda_+}dx (K+V)=T_{\rm
  wall}+(\Lambda_+-\left<x \right>){\cal V}(1), 
\end{eqnarray}
which gives 
a force (pressure) from the SUSY vacuum to the domain wall 
\begin{eqnarray}
  -\frac{d E_{\rm wall}}{d \left<x\right>}={\cal V}(1)>0.
\end{eqnarray}
Moreover, it is natural to require for the relation,  
\begin{eqnarray}
 \int_{\Lambda_-}^{\Lambda_+} dx K
=\int_{\Lambda_-}^{\Lambda_+} d x V- (\Lambda_+-\left<x\right>) {\cal
V}(1), 
\label{eq:modifiedDeric}
\end{eqnarray}
is hold near the domain wall solution. 
When ${\cal V}(1)=0$ holds, the above relation can be  
derived from Derrick's theorem \cite{Vilenkin}. Then, we define 
the tension $T_{\rm wall}$ and a position $\left<x\right>$ of the wall  
in terms of only kinetic terms $K$ without  using a potential $V$ as
\begin{eqnarray}
 T_{\rm wall}\equiv 2 \int_{\Lambda_-}^{\Lambda_+}dx K,\quad 
\left< x\right>\equiv \frac{2}{T_{\rm wall}}\int_{\Lambda_-}^{\Lambda_+}dx\, x K. 
\end{eqnarray}    

The equation (\ref{eq:modifiedDeric}) is enough to estimate  
data $\{T_{\rm wall}, L_{\rm wall}\}$ as the following.
Let us approximate the profile functions $(T,s)$ for the domain wall by piecewise-linear functions as
\begin{eqnarray}
 T=\frac{x}{L_{\rm wall}},  \quad s=1-\frac{x}{L_{\rm wall}}
\quad {\rm for}\quad 0 \le x \le L_{\rm wall},
\end{eqnarray}
and
$(T,s)=(1,0)$ for  $x\ge L_{\rm wall}$ and $(T,s)=(0,1)$ for $x<0$.
By inserting this approximation to Eq.(\ref{eq:modifiedDeric})
we find that the l.h.s (r.h.s) is proportional to $L_{\rm
wall}^{-1} (L_{\rm wall})$. 
Note that the tension of the domain wall can be expressed as
\begin{eqnarray}
T_{\rm wall}=\int_{\Lambda_-}^{\Lambda_+} dx K
+\int_{\Lambda_-}^{\Lambda_+} d x V- (\Lambda_+-\left<x\right>) {\cal
V}(1).
\end{eqnarray}
Minimizing it in terms of $L_{\rm wall}$, we get
\begin{eqnarray}
 L_{\rm wall}&=&\sqrt{
\frac{(A(\lambda)+\epsilon^2)^2}{(A(\lambda)+\epsilon^2) B(\lambda,\epsilon)
-(A(\lambda)-C(\lambda)+ \epsilon^2/2 ){\cal V}(1)}} , \nonumber\\
T_{\rm wall}&=&2\sqrt{\left(A(\lambda)+\epsilon^2\right)B(\lambda,\epsilon^2)
-\left(A(\lambda)- C(\lambda)+\frac{\epsilon^2}{2} \right) {\cal V}(1) },
\end{eqnarray}
where
\begin{eqnarray}
&& A(\lambda)=\int_0^1\frac{dx}{{\cal V}(x)}
=\left\{
\begin{array}{cc}
\displaystyle 1+\frac7{60\lambda}+{\cal O}(\lambda^{-2}) & {\rm for~} \lambda\gg 1 \\ {}\\
\displaystyle \frac{\pi}{4\sqrt{4\lambda-1}}- \frac18 \log(4\lambda-1)
+{\cal O}(1)& {\rm for~} \lambda\sim \frac14
\end{array}\right. ,\nonumber\\
&&B(\lambda,\epsilon)=\frac1{15}\left(8-\frac{233}{42}\frac1{4\lambda}+\frac2{\epsilon^2}\right),\\
&& C(\lambda)=\int_0^1 \frac{x dx }{{\cal V}(x)}=
2\lambda \frac{{\rm arccot}\sqrt{4\lambda-1}}{\sqrt{4\lambda-1}}
=\left\{
\begin{array}{ll}
\displaystyle \frac12+\frac1{12\lambda}+{\cal O}(\lambda^{-2}) 
& {\rm for~} \lambda\gg 1 \\ {} \\
\displaystyle \frac{ \pi}{4\sqrt{4\lambda-1}}
+{\cal O}(1)& {\rm for~} \lambda\sim \frac14
\end{array}\right. . \nonumber
\end{eqnarray}
Here we took $\Lambda_-<0$ and $\Lambda_+>L_{\rm wall}$. 
Especially we find inequality
\begin{eqnarray}
 T_{\rm wall}L_{\rm wall}=2(A(\lambda)+\epsilon^2)< 
2\left(\frac1{{\cal V}(1)}+\epsilon^2\right). 
\end{eqnarray}
We have used this for deriving the inequality in Eq.~(\ref{eq:bound}).

\begin{figure}
\begin{center}
\includegraphics[width=8cm]{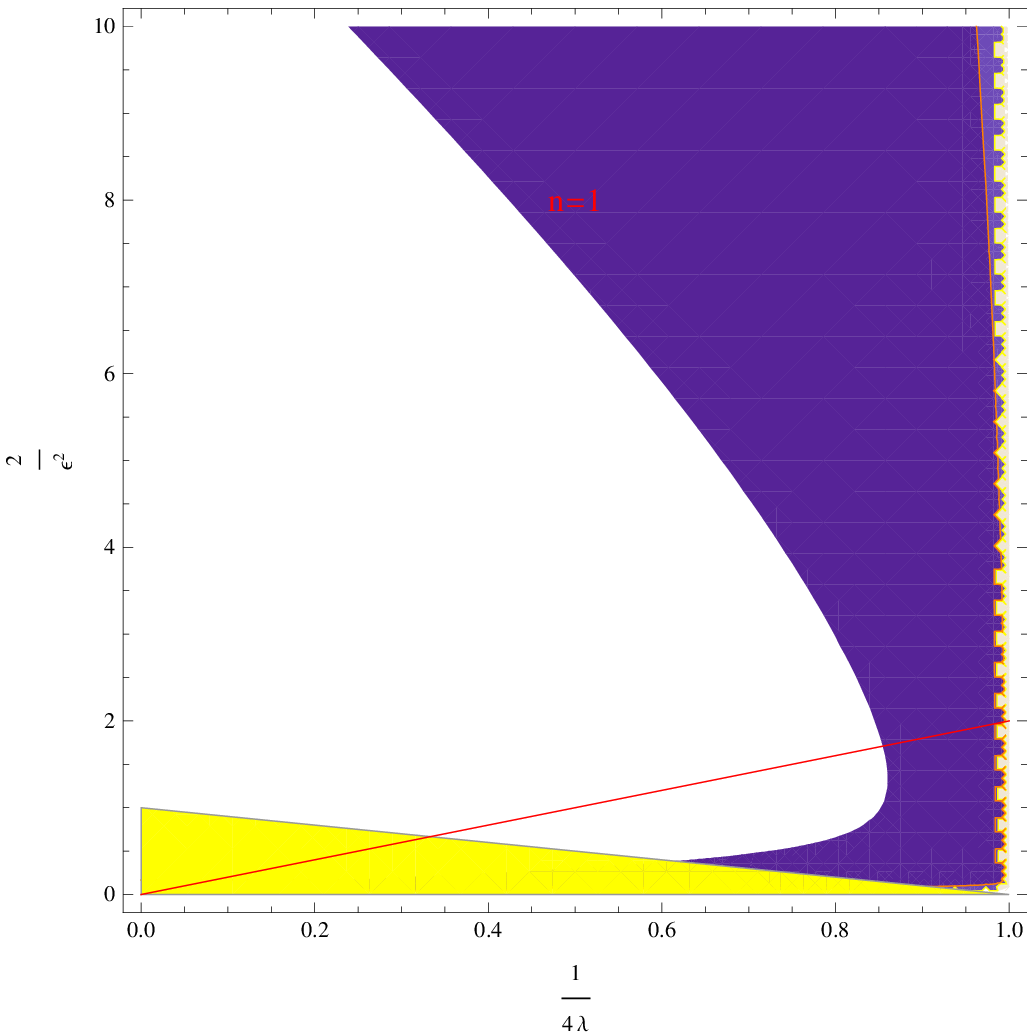}
\includegraphics[width=8cm]{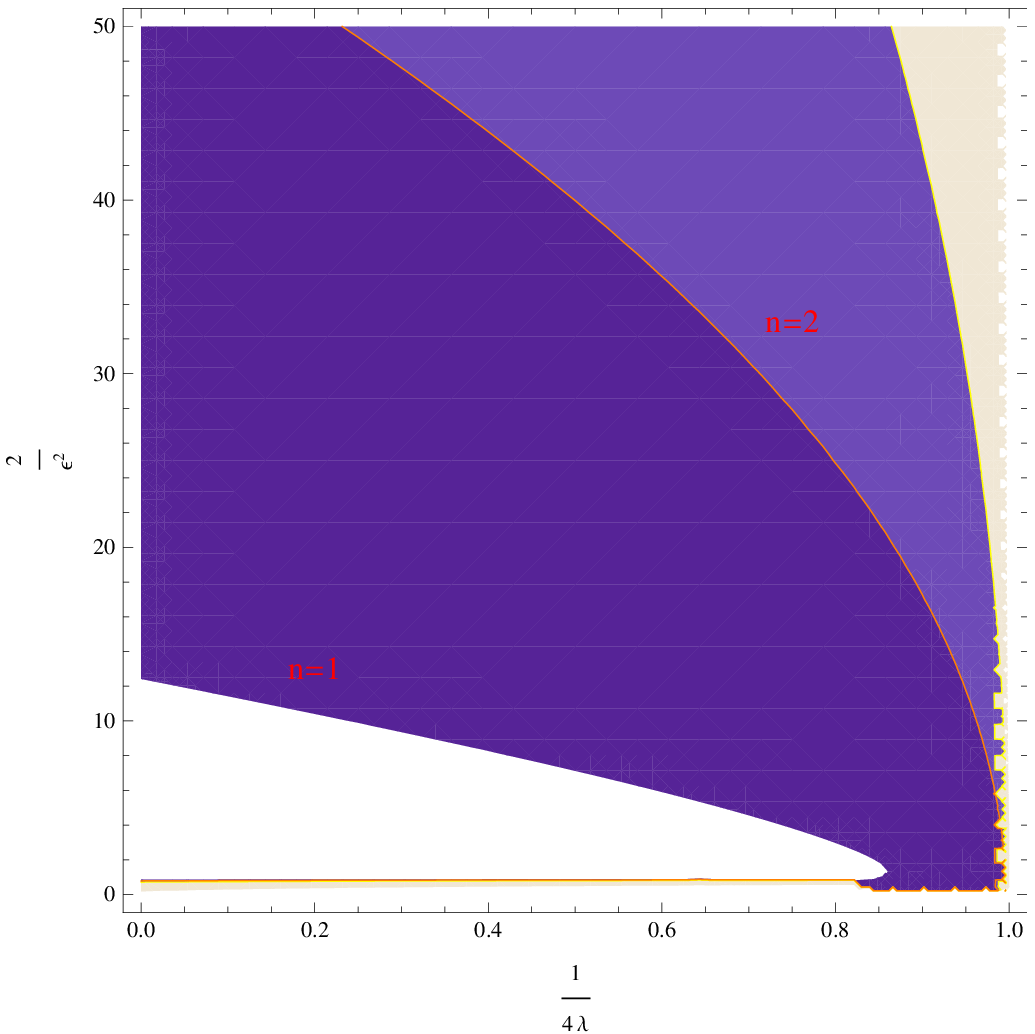}
\caption{\sl Stability of the R-tube with winding number. In the white region, all tube solutions $n\ge 1$ are unstable. In the purple region,R-tube with $n=1$ is stable but others $n \ge 2$ are unstable.  In the light purple region, R-tubes with $n=1,2$ is stable but others $n \ge 3$ are unstable. The region below the red line represents \eqref{redline}. The SUSY breaking vacuum is unstable in the yellow region (see \eqref{eq:VacuumCond}).   }
\label{LinearApp}
\end{center}
\end{figure}

Finally, using the result $T_{\rm wall}$ in the linear approximation, we show the stability condition \eqref{eq:stability1} for 
the R-tube with winding number $n$ in Fig \ref{LinearApp}.

\subsection{Numerical approach}

\subsubsection{Numerical calculation for stability}

In the previous subsection, exploiting linear approximation, we found
the stability condition of the R-tube with winding number $n$. Here,
we try to check the parameter dependence of the stability by numerical
calculation. We adopt a kind of relaxation method to find a
configuration of the R-tube. See Appendix \ref{sec:relaxation} for details. Since we are interested in a parameter region close to borders of the stability of two winding numbers, so we have to treat relatively unstable configuration, which require careful analysis. Because of this complexity, we focus on a couple of examples for the numerical analysis. 

\begin{figure}
\begin{center}
\includegraphics[width=8cm]{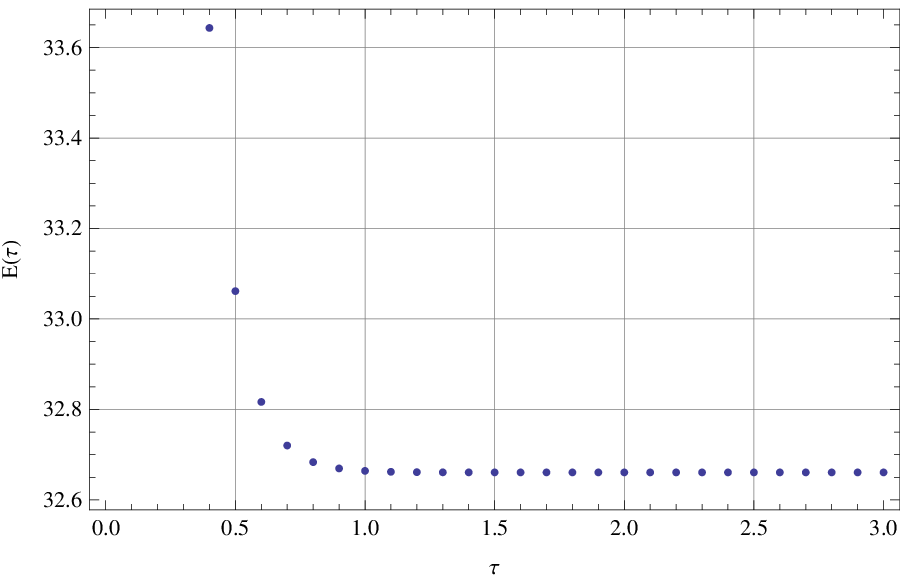}
\includegraphics[width=8cm]{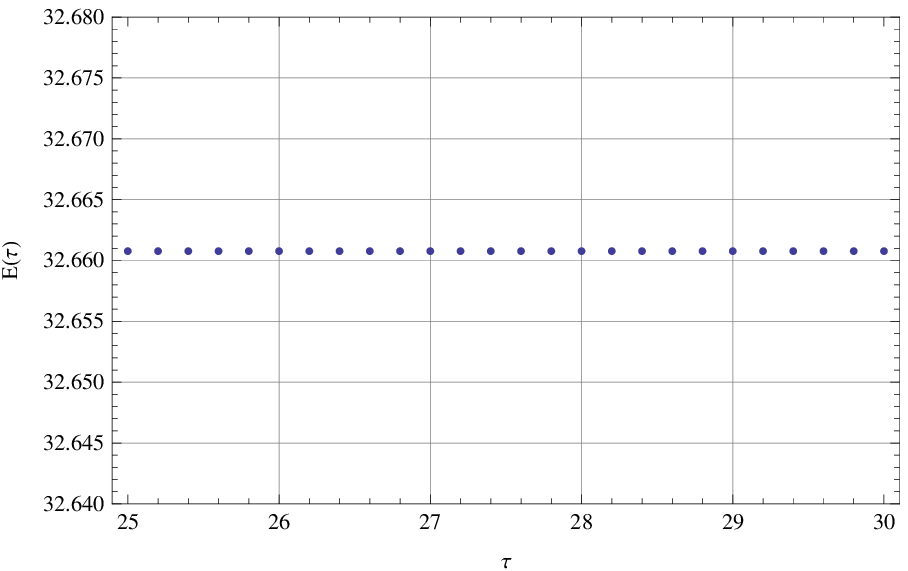}
\caption{\sl Energy against the relaxation time $\tau$. Energy of the configuration with winding number $n=1$ 
for $\epsilon=1$ and $\lambda=27/100$ converges to a constant value. $\tau$ is the relaxation time.}
\label{converge}
\end{center}
\end{figure}

\begin{figure}
\begin{center}
\includegraphics[width=8cm]{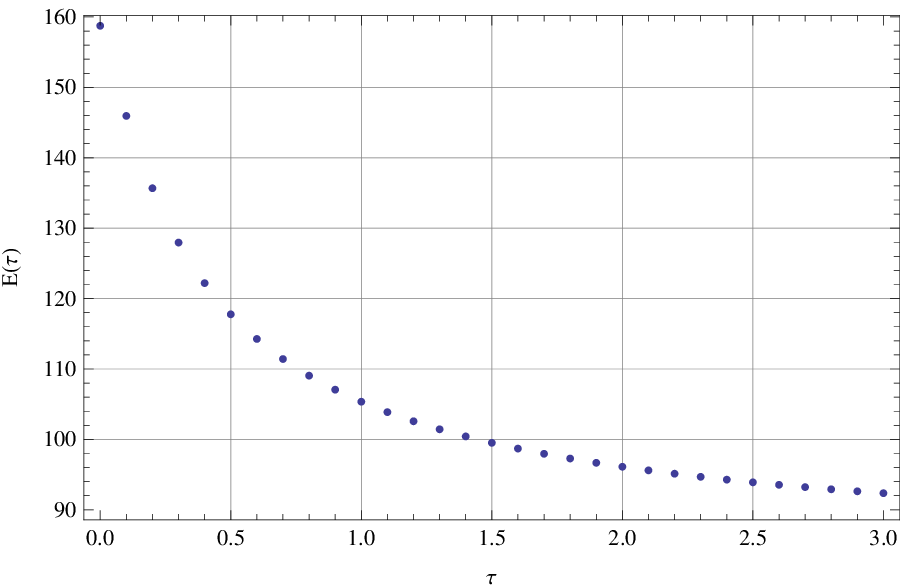}
\includegraphics[width=8cm]{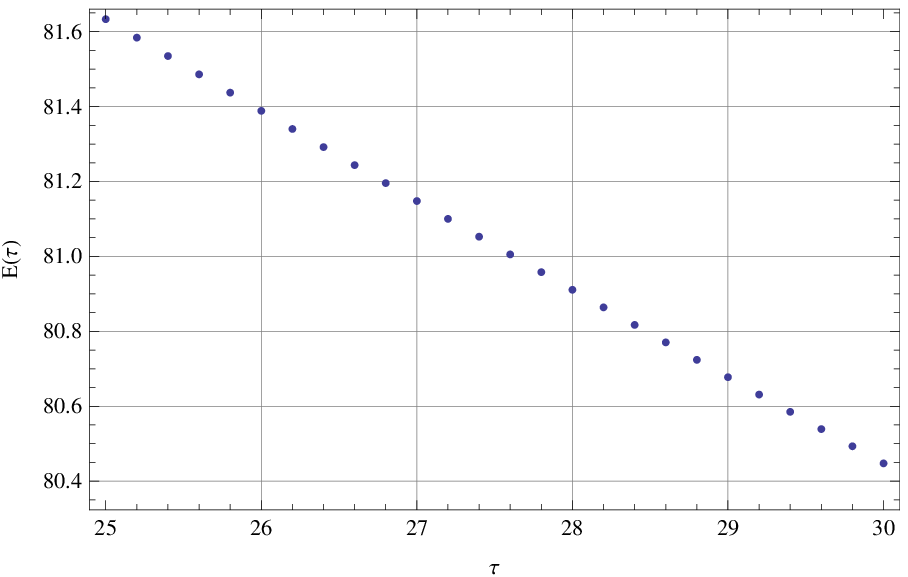}
\caption{\sl Energy against the relaxation time $\tau$. Energy of the configuration with winding number $n=2$ for $\epsilon=1$ and $\lambda=27/100$ decreases linearly at a later relaxation time. }
\label{Higherconverge}
\end{center}
\end{figure}

As a first example, we take a parameter $\epsilon=1$, $\lambda=27/100$
where according to the linear approximation, winding number $n=1$ is
stable but $n=2$ is unstable (see Fig \ref{LinearApp}). Following the
relaxation method, we take an appropriate initial function and finite
relaxation time $\tau$, then we calculate minimum energy
configurations. As we show in Fig \ref{converge} and Fig
\ref{Higherconverge}, energy convergences of the configurations have a
clear difference in two cases. Here we removed a contribution $E_{\rm vev }$ of the vacuum energy
density from the total energy $E$ and calculated the following
dimensionless energy 
\begin{eqnarray}
 E(\tau)\equiv \frac{\lambda}{m^2} \frac{E-E_{\rm vev}}{2\pi \Delta z}
=\frac{\lambda}{m^2} \frac{E}{2\pi \Delta z}-\frac12 {\cal V}(1)
\Lambda^2, \label{eq:dimlessenergy}
\end{eqnarray}
and we take the IR-cutoff of 
the energy as $\Lambda=50$.
The configuration with $n=2$ is monotonically loosing the energy and in a sufficiently late relaxation time $\tau$, the energy decreases as a linear function, which clearly suggests instability of this configuration. On the other hand, as for the configuration with $n=1$, the energy seems to converge to a constant value. This sharp difference nicely matches with the result of the liner approximation. 

However, it is worth  noting that our numerical analysis is done with
a finite precision which is appropriately chosen by reasonable
calculation time.  Thus, beyond our calculation precision, there may
exist an unstable mode which may yield slight energy loss. Thus, as
long as we use a kind of relaxation method with a fixed initial
condition, it may be, in general, hard to conclude complete stability of the configuration. However, even if the small instability exists, the life-time of R-tube can be longer than the decay time of the R-tube originating from an explicit $U(1)_R$ breaking effect. In many phenomenological models, the global R-symmetry is already broken by adding gravity due to the constant term in superpotential. Thus, at a point of the early universe, R-tubes disappear by generating axion domain walls \cite{OurII,Sikivie:1982qv,lyth,nagasawa,chang,hiramatsu}. Therefore, as long as the stability is long enough compared with its lifetime, we can treat the R-tube as a stable solution.

\begin{figure}
\begin{center}
\includegraphics[width=8cm]{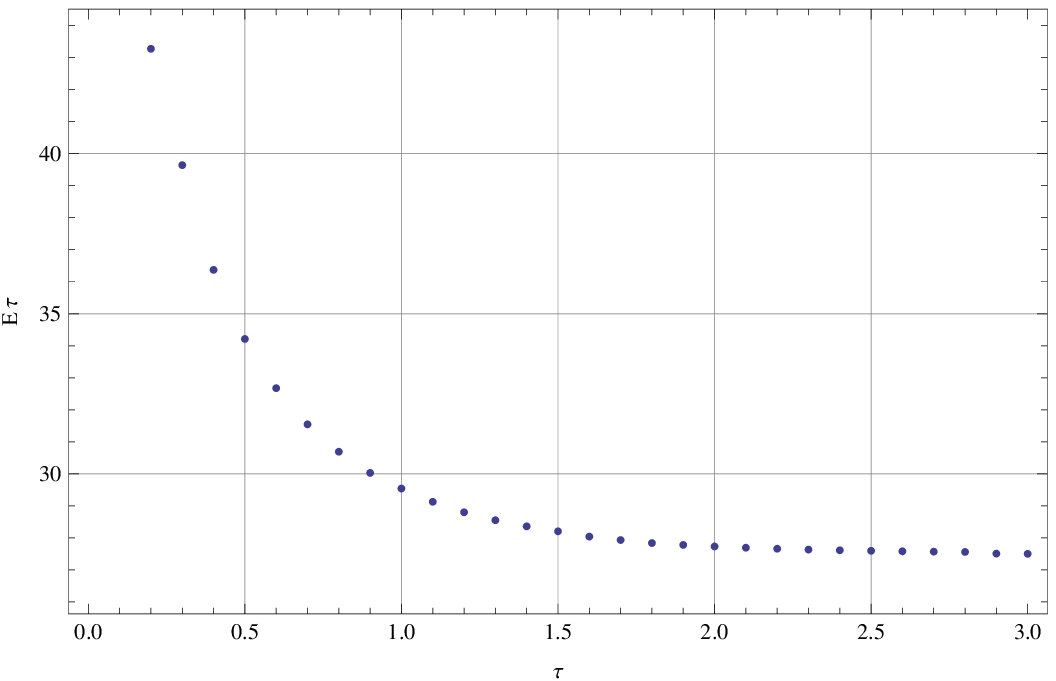}
\includegraphics[width=8cm]{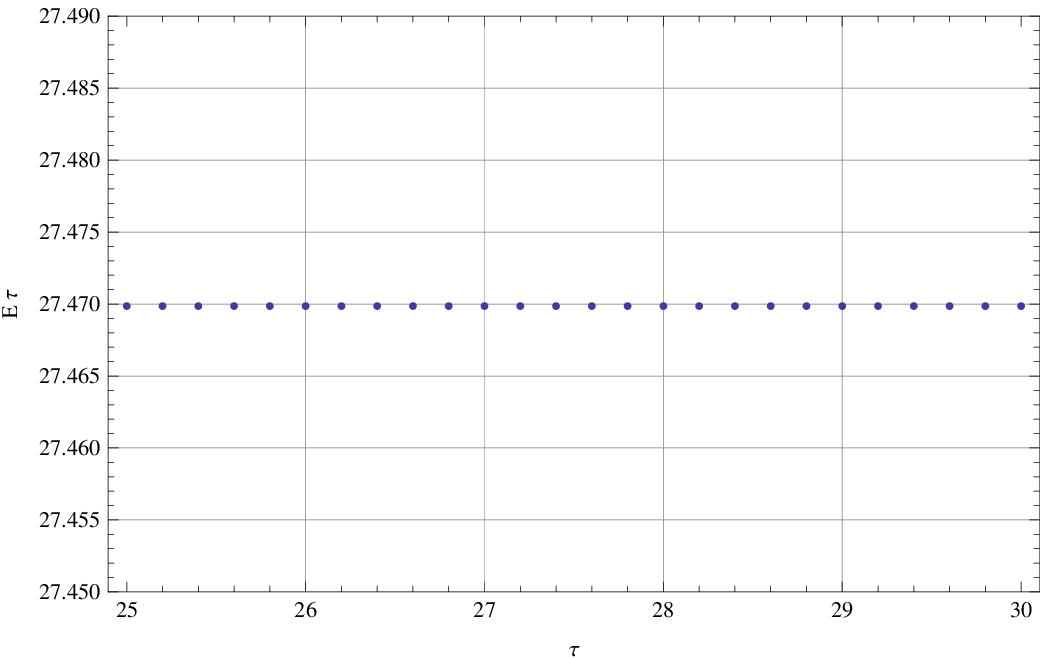}
\caption{\sl Energy against the relaxation time $\tau$. Energy convergence of the configuration with winding number $n=2$ for $\epsilon=1/50$, $\lambda=6/10$. }
\label{converge2}
\end{center}
\end{figure}

As a second example, we choose $\epsilon=1/50$ and
$\lambda=6/10$. Small $\epsilon$ is favorable in model building,
partially because the longevity of the false vacuum. So from
phenomenological point of view, it is important to study a
configuration of R-tube with small winding number in this parameter
region. Again, using the relaxation method, we numerically calculate
the energy convergence of two cases, $n=1$ and $n=2$. As shown in Fig
\ref{converge2}, the total energy of R-tube with $n=2$ converges to a
constant value. Thus, within our calculation accuracy, the
configuration looks stable. Also, the confituration with $n=1$ is
similarly stable. With these numerical results and linear
approximation shown in the previous subsection, it may be plausible that in small $\epsilon$ region R-tubes are relatively stable and the roll-over process does not occur.

\subsubsection{Effective potential for light mode}

As emphasized above, there may exist a very light mode which may cause
instability of a configuration. Although treatment of such light modes
in the relaxation method is not an easy task, but we would like to
propose a method to uncover the existence of a light mode. The most
interesting mode is a fluctuation of the size of the R-tube. Generally speaking a zero mode (moduli) around a solution is frozen in the relaxation method,
and a light mode of which dependence in the total energy is quite
small, seems to be almost frozen even if it exists. 
To detect such light mode and search the true stable solution,
we need to take a lot of different initial conditions for the relaxation
method. To be concrete, we show initial conditions for the fields $T,s$,
\footnote{
We set Dirichlet condition $f(0)=0$ and Neumann condition $h'(0)=0$ for
$T$ and $s$ for respectively.
We need to be sensitive for consistency between initial conditions and
boundary conditions at $\rho=0$.}
\begin{eqnarray}
&& f(\rho)=\frac{1+\tanh(2(\rho-\rho_0))}2 \tanh(\rho), \\ \nonumber 
 && h(\rho)=\frac{1-\tanh(2(\rho-\rho_0))}2
\frac{1+\tanh(2(\rho+\rho_0))}2.
\end{eqnarray}
With various values for $\rho_0$ which  roughly indicates a transverse
size of R-tube, we calculate minimum energy
configurations with finite relaxation time. 
For any value of $\rho_0$, the energy converges like 
Figs. \ref{converge} and \ref{converge2} for those values of $n$,
$\epsilon$ 
and $\lambda$.
However, 
final configurations can have small differences of the energy and the
tube-size. To represent the size of the tube, it would be useful to
introduce the 
following definition similar to \eqref{eq:RstringSize2},
\begin{eqnarray}
 R_T\equiv \frac{\int_0^\infty d\rho \rho^2 K_T}{\int_0^\infty d\rho \rho
  K_T},\quad  R_s\equiv \frac{\int_0^\infty d\rho \rho^2 s'(\rho)^2}{\int
  _0^\infty d\rho \rho s'(\rho)^2}. \label{ini}
\end{eqnarray}
Here we defined two sizes of the R-tube, $R_T$ and $R_s$. A reason for introducing two sizes can be seen in a discrepancy between the linear approximation in section \ref{Linear} and numerical results shown below. As has mentioned in section  \ref{sec:R-string}, these quantities do not include the cut-off dependence and are well-defined candidates for the size of the tube.

\begin{figure}
\begin{center}
\includegraphics[width=7cm]{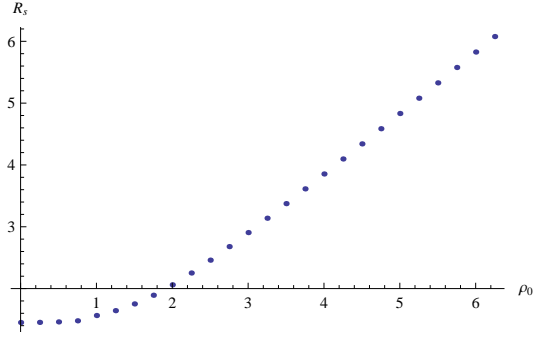}
\caption{\sl One to one correspondence between the initial condition $\rho_0$ and the size $R_s$ for $\epsilon =1$, $\lambda=27/100$.}
\label{rho0vsRs}
\end{center}
\end{figure}

As an example, we take $\lambda=27/100$, $\epsilon=1$. Varying the
initial position $\rho_0$, we calculate the minimum energy configuration with finite relaxation time. First of all, we show a correspondence between the initial condition $\rho_0$ and the size $R_s$ in Fig \ref{rho0vsRs}. $R_s$ is evaluated with a converged configuration. Since it is one-to-one correspondence, varying the initial condition $\rho_0$ represents varying the size of the R-tube. 

\begin{figure}
\begin{center}
\includegraphics[width=8cm]{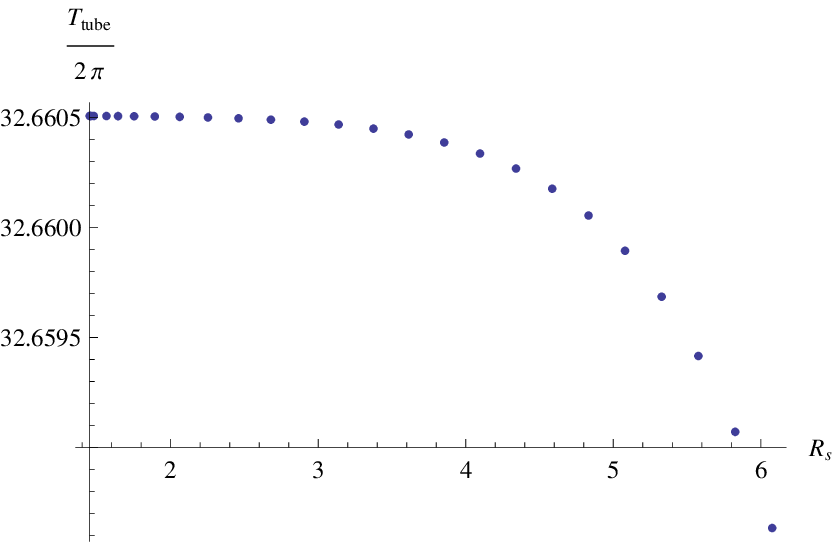}
\includegraphics[width=8cm]{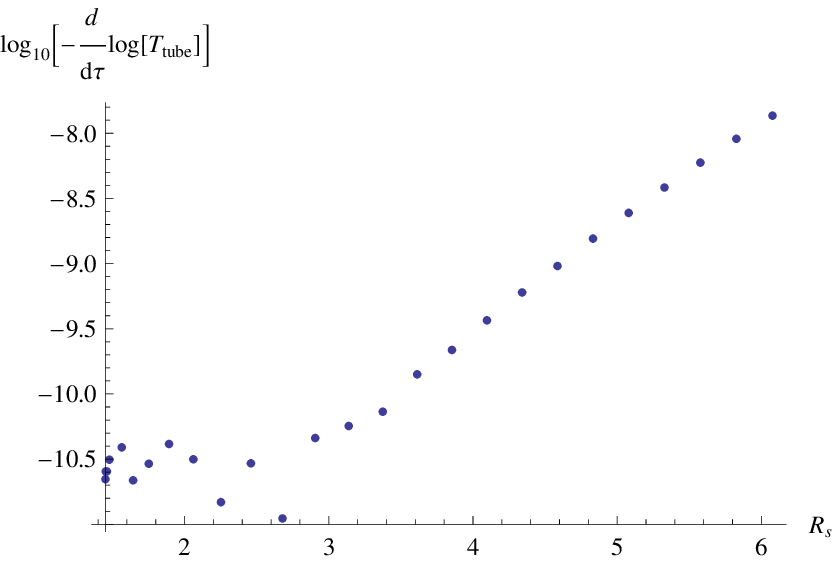}
\caption{\sl Monotonically decreasing potential of $R_s$ with $\lambda=27/100,\epsilon=1$(the left
 panel). $T_{\rm tube}\equiv 2\pi E(\tau=50)$. Gradients in the energy with respects to $\tau$ at $\tau=50$(the
 right panel). Finite gradients indicate instability of configurations and noises for small $R_s$ imply the limit bound of precision of calculations.}
\label{fig:potential1}
\end{center}
\end{figure}

\begin{figure}
\begin{center}
\includegraphics[width=8cm]{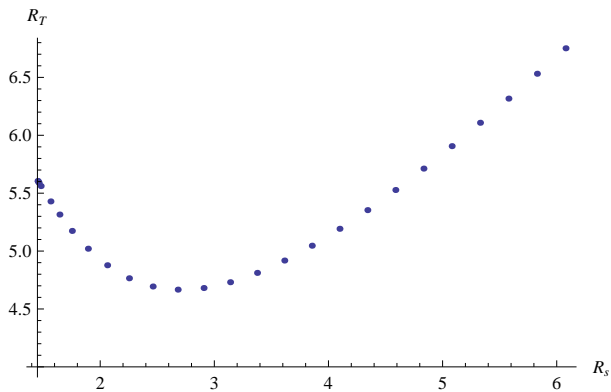}
\caption{\sl Relation between the two transverse sizes $R_T,
 R_s$ with $\lambda=27/100,\epsilon=1$. 
$R_T$ is decreasing for small $R_s$ whereas $R_T$  for large $R_s$
 is proportional to $R_s$. Therefore $R_T$ is not  a good quantity for
 parametrizing the potential.}
\label{fig:RTvsRs}
\end{center}
\end{figure}

Now let us show the low energy effective potential for the fluctuation mode of the size. We plot the energy \eqref{eq:dimlessenergy} at the relaxation time $\tau=50$, which we will denote as $T_{\rm tube}\equiv 2\pi E(\tau=50)$, with respect to the position of $R_s$ in Fig.\ref{fig:potential1}. Surprisingly, we observe 
a monotonically decreasing potential in terms of $R_s$  for the model with
 $\lambda=27/100,\epsilon=1$ 
which we explained. A large plateau with tiny gradient in
Fig.\ref{fig:potential1} is consistent with Fig.\ref{converge} where 
$R_s$ can be almost regarded as a massless mode. 
Fig.\ref{fig:potential1} implies that the R-tube in this case is unstable
and will expand to the infinity. This clearly suggests that the border line of instability of minimum winding R-tube shown in Fig \ref{LinearApp} does not match with the numerical analysis above. We find that for small $R_s$ 
the two sizes behave differently as shown in Fig.\ref{fig:RTvsRs}
and this fact tells us why the estimation (\ref{eq:toymodel}) with
a single size shown in section \ref{Linear} does not
work for small $R_s$. It would be interesting to study two-scale linear approximation for better approximation.

Here, we proposed a way to analyze the low energy effective theory for
a very light mode corresponding to the fluctuation of the tube-size and
checked the instability of the mode.
However, this light mode is significantly affected by various
corrections such as thermal effects, supergravity effects 
and quantum corrections.
We would study these corrections elsewhere.     

\section{Bamboo solution: Tube junction}

In the previous section, we showed that an R-string can form a
tube-like domain wall with winding number and inside of the wall is in
the SUSY preserving vacuum. It would be wonderful if we could extract any evidences of the existence of the SUSY vacuum from outside of the tube\footnote{The global R-symmetry is explicitly broken when gravity is coupled to the theory. In this case, when the Hubble parameter $H$ becomes the mass scale of the $R$-axion, a domain wall interpolating the strings is generated and string and walls disappear \cite{OurII,Sikivie:1982qv,lyth,nagasawa,chang,hiramatsu}. Thus, one cannot observe a global R-string in the present age. However if we replace the $R$ symmetry with another local $U(1)$ symmetry, then a similar tube-like solution existing in the present age can be generated.}. Toward this goal, in this section we demonstrate that quantum number in the SUSY vacua significantly affects the shape and the tension of the string. Concretely, we construct a junction of the R-tube.
To the best of our knowledge, this kind of soliton has not been known so far.
In order to demonstrate an explicit solution, let us again take the two-scalar model in Eq.~(\ref{eq:model_sol3}).
As shown in Eq.~(\ref{eq:ansataz_sol3}), reflecting the $\mathbb{Z}_2$ symmetry of the model,
there are two different R-tubes\footnote{This is reminiscent of the monopole junction of two cosmic strings studied in \cite{HashimotoTong}. It would be interesting to study further our junction in light of this similarity.}. The one has $\gamma = +1$ and the other has $\gamma = -1$.
The skin of the R-tube, namely the profile of $T$ field, is independent of the choice of $\gamma$, so that
one can naturally imagine that the two R-tubes with different $\gamma$ can be smoothly connected.
The junction of the R-tubes is a domain wall which interpolates two different SUSY vacua inside the R-tube.
We call this junction the R-bamboo.

\begin{figure}
\begin{center}
\includegraphics[width=14cm]{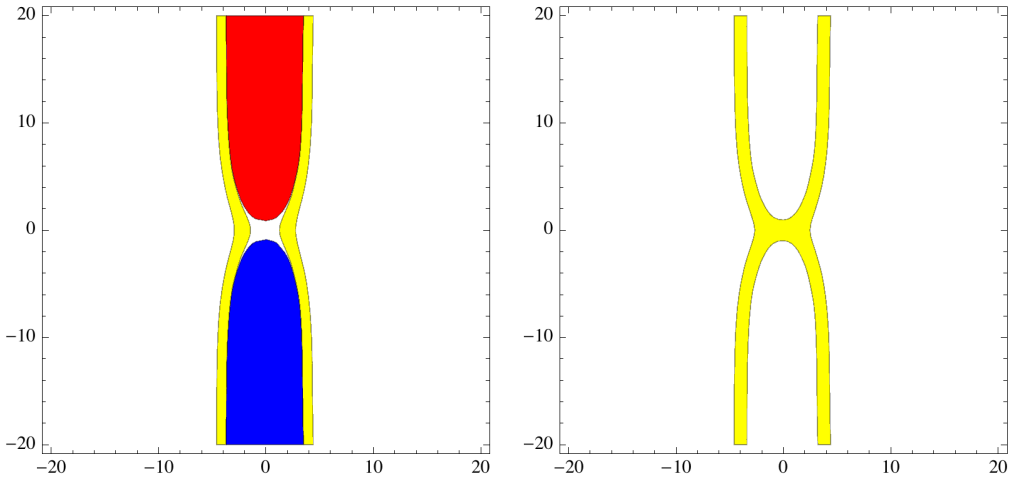}
\caption{\sl The bamboo soliton (A slice at the center): Junction of the R-tubes with $\gamma =+1$ and $\gamma=-1$. 
In the left panel, the red region is where $s > 1/2$ while the blue is the region where $s < -1/2$,
and the yellow stands for the region where $7/20<|T|<12/20$. The right panel shows the potential density isosurface
with which one can clearly recognize the domain wall inside the tube.}
\label{fig:bamboo_slice}
\end{center}
\end{figure}
\begin{figure}
\begin{center}
\includegraphics[height=7.5cm]{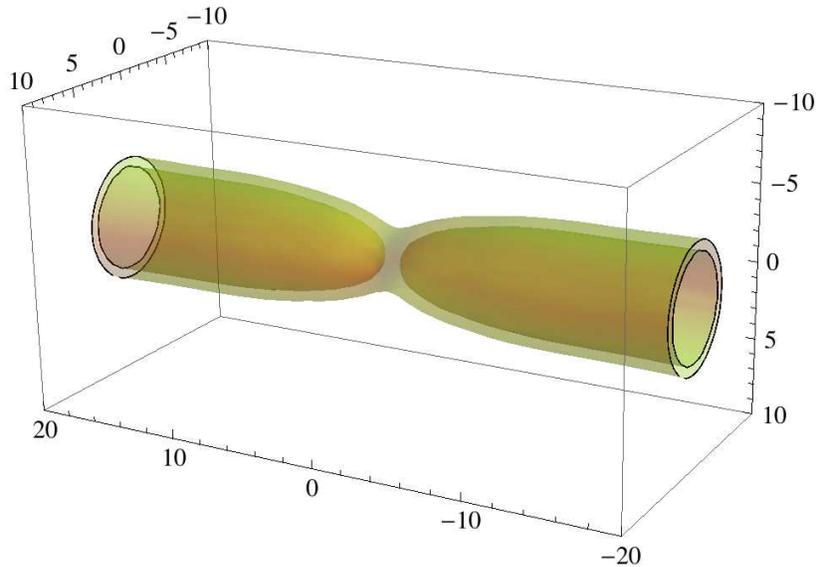}
\caption{\sl 3D plot of the Bamboo soliton for $\epsilon=1$, $\lambda=27/100$.}
\label{fig:bamboo_3d}
\end{center}
\end{figure}
A numerical solution for $\epsilon=1$, $\lambda=27/100$ is shown in Figs.~\ref{fig:bamboo_slice} and \ref{fig:bamboo_3d}.
Far away from the domain wall along the tube, the solution asymptotically goes to the R-tube solution.
At the junction, the transverse size of the tube becomes smaller since the domain wall pulls the
tube toward its inside, see Fig.~\ref{fig:bamboo_slice}.

This R-bamboo solution may be created when two R-tubes collide. If the two tubes 
are different kind, the domain wall must be created at the junction of the two tubes.
At the same time, the anti R-bamboo may be created.
This is very similar phenomenon to monopole and anti-monopole creation associated with
the non-Abelian string reconnection.
This is interesting issue but is beyond the scope of this paper, so we leave it as a future work.

Finally, it is worthy to note that stability of bamboo configuration is not guaranteed by our numerical approach. As emphasized in the previous section, our analysis is a kind of relaxation method. Actually, if there is no domain wall inside the R-tube, the configuration seems to have a small instability shown in Fig \ref{fig:potential1}. However, in the bamboo configuration, the existence of the domain wall increases the stability because the energy of the domain wall is proportional to $R^2$. Thus, if the number of the domain walls is large, the bamboo configuration would be stabilized enough. In principle, the same argument done in Fig \ref{fig:potential1} would be applicable, however numerical calculation becomes quite involved because one needs the two-dimensional relaxation method.


\section{Conclusion}


In this paper, we have demonstrated a fascinating role of a cosmic R-string/R-tube by using two toy models with spontaneous R-symmetry breaking. The first example shown in section \ref{sec:R-string} is a single field model. The model can be regarded as a toy model of one of O'Raifeartaigh models studied in \cite{Shih:2007av} in which pseudo-moduli space is stable everywhere. In this model, string-like defect generated by the Kibble-Zurek mechanism is stable and very close to the known global strings. Winding number dependence of the size of the string is linear in $n$. On the other hand, in the two-field model shown in section \ref{R-tubeInstability}, a string-like object is a tube-like domain wall interpolating a false vacuum and a true SUSY vacuum. One specific feature of two-field model is existence of a tachyonic direction at R-symmetry restoring point $X=0$. Because of this, the core of the string is not stable under fluctuation toward the tachyonic direction. Thus, inside the core in which R-symmetry is restored, can be filled out by the true vacuum through the tachyonic direction. Naively, one may think that such an R-tube is unstable. However, as we shown in the main text numerically, there exist metastable R-tube solutions for certain parameters. An interesting property of such R-tube is the winding number dependence of the size. By the linear approximation, we estimated the dependence and found that its dependence is $n^2$ rather than $n$. Therefore, there is a tendency that tube configuration with larger winding number is more unstable. Numerically, we checked the higher winding instability at a sample parameter $\epsilon=1$ and $\lambda=27/100$: We showed that the configuration with winding number $n=2$ is unstable and the total energy decreases monotonically.

If an unstable R-tube is created by the Kibble-Zurek mechanism, it
rapidly expands and our universe will be completely filled by the true
vacuum. This process gives  constraints for models building. However,
it is worthy to emphasize that such roll-over process can be protected by D-term contribution or thermal potential. As is demonstrated in \cite{NakaiOokouchi,AzeyanagiKobayashi,Azeyanagi:2012pc}, when a D-term contribution cannot be negligible, it can lift the tachyonic direction and stabilize the pseudo-moduli space. In such models, the roll-over process does not occur. Also, when the amplitude of (tachyonic) messenger mass at the origin is sufficiently smaller than that of R-symmetry breaking field, the vacuum selection is successfully realized by exploiting the thermal potential or Hubble induced mass. If such thermal potential keeps lifting the tachyonic direction until the time of the R-string decay \cite{OurII,Sikivie:1982qv,lyth,nagasawa,chang,hiramatsu}, the roll-over process can be successfully circumvented. However, even if such an early stage scenario is assumed, there are sever cosmological constraints on R-axion density as studied in \cite{OurII}.

Finally we comment on our numerical analysis done in section \ref{R-tubeInstability}. Since we adopted a kind of relaxation method with a fixed initial condition to find energetically minimum configurations, it is not easy to conclude the stability of the configuration sharply. There may exist a very light mode which does not change the energy significantly. To uncover the existence  of the very light mode, we proposed a method for studying the low energy effective potential for such light mode. By changing the initial conditions and observing converged configurations, we can estimate the effective potential. Using this method, at the parameter $\epsilon=1$ $\lambda=27/100$, we find emergence of a light model and find a large plateau in the effective potential. It would be important to apply the method to a wide range of parameter space and study borders of instabilities of configurations with various winding numbers numerically. This is beyond the our scope, so we will leave it as a future work.

\section*{Acknowledgement}

The authors would like to thank T.~Hiramatsu and A.~Ogasahara 
for useful discussions. KK would like to thank Kyoto University for their hospitality where this work was at the early stage. 
The work of M.~E. is supported by
Grant-in-Aid for Scientific Research from the Ministry of Education, Culture, Sports, Science and Technology, Japan (No. 23740226)
and Japan Society for the Promotion of Science (JSPS) and Academy of Sciences of the Czech Republic (ASCR) 
under the Japan - Czech Republic Research Cooperative Program.
TK is supported in part by the Grant-in-Aid for the Global COE 
Program "The Next Generation of Physics, Spun from Universality and 
Emergence" and the JSPS Grant-in-Aid for Scientific Research
(A) No. 22244030 from the Ministry of Education, Culture,Sports, Science and 
Technology of Japan. YO's research is supported by The Hakubi Center for Advanced Research, Kyoto University.

\appendix

\section{Relaxation method}\label{sec:relaxation}
In this section we review a kind of relaxation methods applied to
constructing numerical solutions in this paper. 
Let us consider the following general 
Lagrangian for scalars $\phi^\alpha$ in $\mathbb R^{d+1}$ with
coordinates $\{x^\mu\}=\{t,x^i\}$,
\begin{eqnarray}
{\cal L}=\frac12 g_{\alpha \beta}\partial_\mu \phi^\alpha \partial^\mu \phi^\beta-V(\phi),
\end{eqnarray}
where $g_{\alpha \beta}$ is the metric for the field space. 
Our goal is to find an numerical static solution
$\phi^\alpha(t,x^i)=\phi_{\rm sol}^\alpha(x^i)$ for this system.
The shooting method is a good strategy for a system with a single scalar fields 
and a single spatial coordinate $x^1$, but only in that case.   
For systems with multi fields in higher dimensions, the shooting method
does not work very well and we need the relaxation method explained bellow.  
Let us introduce a `relaxation time' $\tau$ instead of the real time $t$
and suppose that 
$\phi^\alpha$ depend on $\tau$ as $\phi^\alpha(\tau, x^i)$.  
Then $\tau$ dependence of $\phi^\alpha$ is defined by, 
\begin{eqnarray}
 \partial_i\frac{\delta \cal L}{\delta \partial_i \phi^\alpha}-
\frac{\delta \cal L}{\delta \phi^\alpha}=-  g_{\alpha \beta}
\frac{\partial \phi^\beta}{\partial \tau}, \label{eq:relaxation}
\end{eqnarray}
with Neumann condition at the boundary of the region $\Sigma\subset
\mathbb R^d$
\begin{eqnarray}
 {\bf n}^i \partial_i\phi^\alpha  \Big|_{x^i\in \partial \Sigma}=0.
\end{eqnarray}
The added term in the r.h.s of Eq.(\ref{eq:relaxation}) i the so-called 
friction term.
Actually, due to this term  we can show that 
the ordinary total energy $E$ with integral region $\Sigma$  
is no longer constant but a monotonic decreasing function $E=E(\tau)$ 
of $\tau$  as
\begin{eqnarray}
 \frac{d E(\tau)}{d \tau}=-\int_\Sigma d^d x g_{\alpha\beta}
\frac{\partial \phi^\alpha}{\partial \tau}
\frac{\partial \phi^\beta}{\partial \tau} <0.
\end{eqnarray}
If we observe the energy converges, we find a solution
$\phi^\alpha_{\rm sol}$ as
\begin{eqnarray}
 \lim_{t\to \infty} E(\tau)=E_{\rm sol} 
\quad \Leftrightarrow \quad \lim_{\tau\to \infty}\frac{\partial
\phi^\alpha(\tau,x^i)}{\partial \tau}=0 \quad \Leftrightarrow \quad 
\lim_{\tau\to \infty} \phi^\alpha(\tau,x^i)=\phi^\alpha_{\rm sol}(x^i).
\end{eqnarray}
If the energy has the global minimum, therefore, 
we can obtain, at least, one static solution 
$\phi^\alpha_{\rm sol}$ by using the relaxation method.  
If there exists multiple local minima of the energy,
an initial condition of $\phi^\alpha$ chooses one of them.
 
\begin{figure}
\begin{center}
\includegraphics[height=4cm]{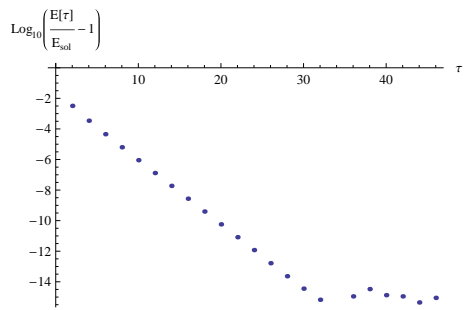}
\caption{\sl Typical behavior of an energy in the relaxation method. Here we
 get accuracy of $10^{-7.5}$ for a solution. Calculations after 
$\tau \approx 32$ turns out to be meaningless. }
\label{fig:EnergyConvergence}
\end{center}
\end{figure}

In actual numerical computation we need a cutoff of the relaxation time 
at $\tau=\tau_{\rm fin}$ and we regard $\phi^\alpha(\tau_{\rm fin},x^i
)$  as a solution.  A relation between $\tau_{\rm fin}$ 
and precision of a solution $\phi^\alpha=\phi^\alpha(\tau_{\rm
fin},x^i)$ can be discussed in the following.
If deviations of $\phi^\alpha$ 
from a solution $\phi^\alpha_{\rm sol}$ 
are  efficiently small, they can be expanded as 
\begin{eqnarray}
 \phi^\alpha-\phi^\alpha_{\rm sol}\approx \sum_{n=1}^\infty f_n(x^i) a_n(\tau),
\end{eqnarray}
with  profile functions $f_n(x^i)$ for massive modes of mass $m_n$
around the configuration $\phi_{\rm sol}^\alpha$.
Eq.(\ref{eq:relaxation}) gives development of coefficients
$a_n(\tau)$  as
\begin{eqnarray}
a_n(\tau)= a_n^0 e^{-m_n^2 \tau}, 
\end{eqnarray} 
and  behavior of the energy is controlled by the lightest mass $m_1$ as
\begin{eqnarray}
 E(\tau)\approx E_{\rm sol}+A e^{-2 m_1^2\tau}.
\end{eqnarray}
Here a constant $A \in \mathbb R_{>0}$ depends on an initial condition 
we took and is assumed to be the same order as $E_{\rm sol}$.
To get precision $10^{-p}$, therefore, we have to take a time $\tau_{\rm fin}$
for this relaxation method as
\begin{eqnarray}
 \tau_{\rm fin}> \frac{p}{m_1^2} \ln 10,
\end{eqnarray}
where $m_1$ can be roughly guessed by a typical mass scale of the system.
With large $\tau$, 
we sometimes observe a random behavior of $E(\tau)$ 
which is a signal of the limit bound of machine precision. 
See Fig.\ref{fig:EnergyConvergence} for an example.   

If a configuration of $\phi^\alpha$ is accidentally near to a saddle point,
we also observe an exponential decay of $E(\tau)$, but after that 
it collapses like a waterfall as
\begin{eqnarray}
 E(\tau)\approx E_{\rm saddle}+ A e^{-2m^2 \tau}-\tilde A e^{2|\tilde m^2|\, \tau}
\end{eqnarray}
with a tachyonic mass $\tilde m^2=-|\tilde m^2|$. 
Therefore  an exponential behavior of the energy do not always guarantee
that a stable solution  is obtained. 
Taking multi initial conditions of $\phi^\alpha$ can avoid 
this technical error.   
%
%

\end{document}